\documentclass[multphys,vecphys]{svmult}

\usepackage{makeidx}         
\usepackage{graphicx}        
\usepackage{multicol}        
\usepackage[bottom]{footmisc}

\makeindex             

\begin{document}

\title*{The parallel lives of supermassive black holes and their host
  galaxies}
\titlerunning{Parallel lives of supermassive black holes
  and galaxies}

\author{Andrea Merloni$^{1}$, Gregory Rudnick$^{2}$, Tiziana Di Matteo$^3$}
\institute{$^{1}$MPI f\"ur Astrophysik,
  K.Schwarzschildstr.1, 85741 Garching, Germany\\
$^{2}$NOAO, North Cherry Ave.,
Tucson, AZ 85721, USA\\
$^{3}$Carnegie Mellon University, Department of Physics, 5000 Forbes
  Ave., Pittsburgh, PA 15213, USA\\}

\authorrunning{A. Merloni, G. Rudnick, T. Di Matteo}

\maketitle

\begin{abstract}
   We compare all the available observational data on the redshift
  evolution of the total stellar mass and star formation rate density
  in the Universe with the mass and accretion rate density evolution
  of supermassive black holes, estimated from the hard X-ray
  selected luminosity function of quasars and active galactic nuclei
  (AGN). We find that on average black hole mass must have been higher
  at higher redshift for given
  spheroid stellar mass. Moreover, 
we find negative redshift evolution of the disk/irregulars to spheroid
  mass ratio. The total accretion efficiency is 
constrained to be between 0.06 and 0.12,
  depending on the exact value of the local SMBH mass density, and on
  the critical accretion rate below which radiatively inefficient
  accretion may take place. 
\end{abstract}

\section{Introduction}
\label{mer_sec:intro}
Observational
evidence indicates that the mass of supermassive black holes (SMBH) is
correlated with the luminosity (Marconi \& Hunt, 2003
and references therein) and velocity
dispersion (Tremaine et al., 2002 and references
therein) of the host spheroids, 
suggesting that the process that leads to
the formation of galaxies must be intimately linked to the
growth of the central SMBH. Studying low redshift AGN, 
Heckman et al. (2004) have shown that
not only does star formation directly trace AGN activity, but also  
that the sites of SMBH growth must have shifted to smaller masses at
lower redshift, thus mimicking the ``cosmic downsizing'' scenario
first put forward to describe galaxy evolution by Cowie et al. (1996).
Such a scenario has recently received many independent confirmations,
both for the evolution of SMBH as traced directly by X-ray and radio
luminosity functions (LF) of AGN (Marconi et al. 2004; Merloni 2004; Hasinger et al. 2005), and for that of
star forming galaxies, thanks to large surveys such as 
SDSS, GDDS, COMBO-17, GOODS, etc. 
(see, e.g. Heavens et al. 2004; Juneau et al. 2005; 
P\'erez-Gonz\'alez et al. 2005; Feulner et al. 2005). 

Following Merloni, Rudnick and Di
Matteo (2004; MRD04), here we discuss a {\it quantitative} 
approach to the study of the posited link between star
formation and SMBH growth, based on a detailed comparison of the redshift
evolution of integral quantities, such as the total stellar mass, black
hole mass and star formation rate densities.

\begin{figure}
\centering
\begin{tabular}{c}
\includegraphics[height=5.cm,width=10.cm]{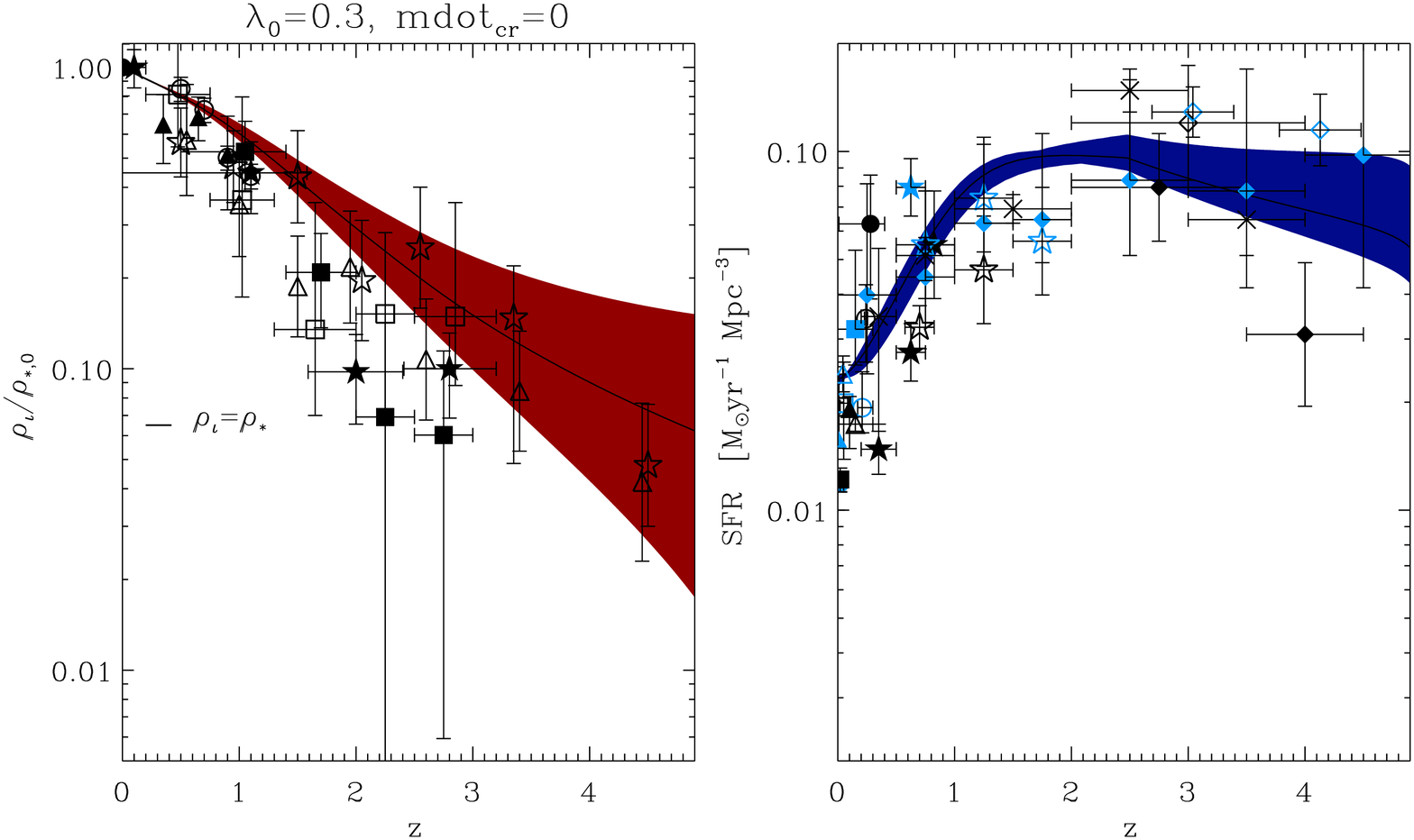} \\
\includegraphics[height=4.46cm,width=10.cm]{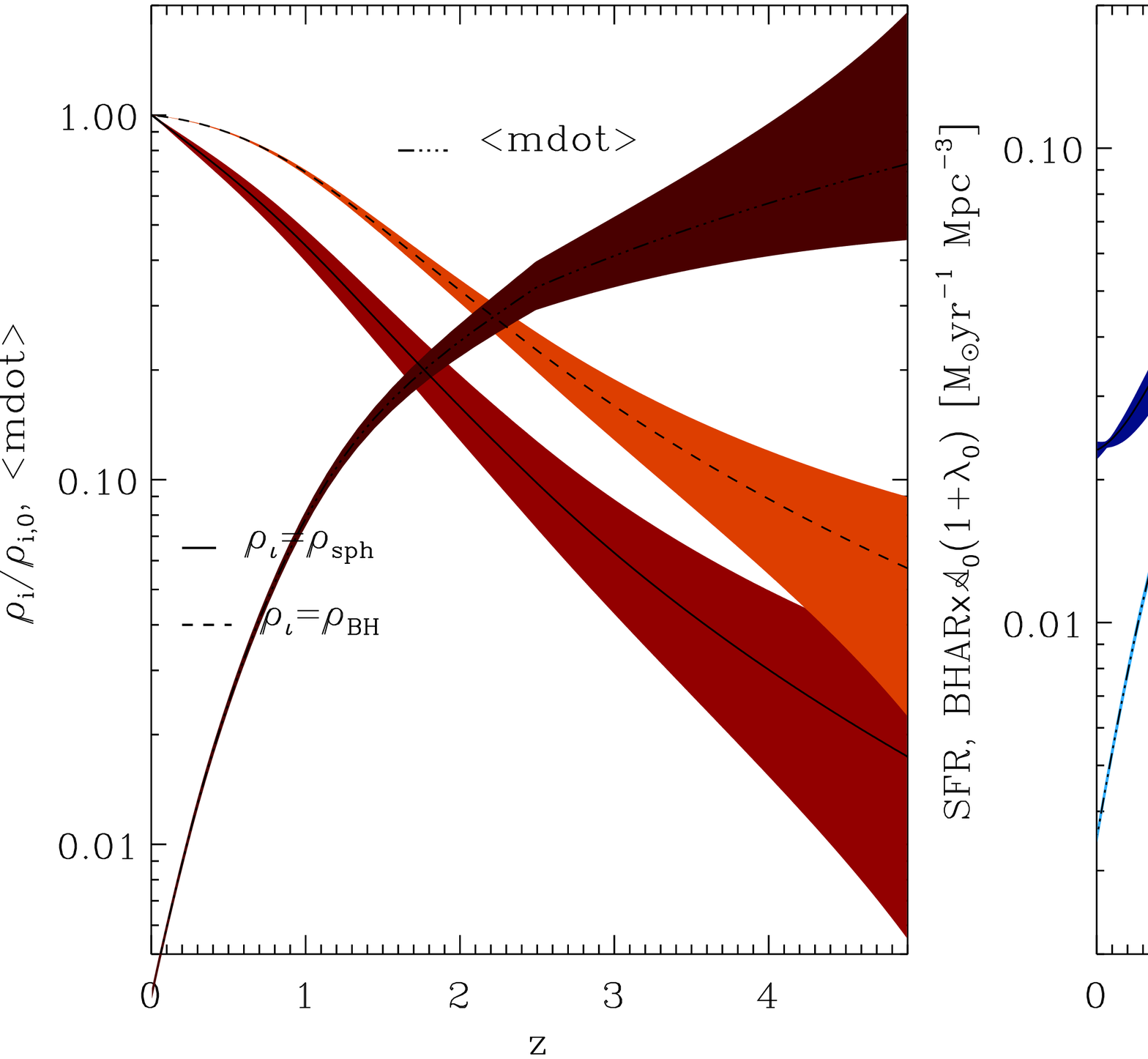} 
\vspace{-0.3cm}
\end{tabular}
\caption{The upper left panel shows the evolution of the stellar 
mass density as a function of redshift, where the density
is given as a ratio to the local value, $\rho_{*,0} = 5.6 \times
10^8 M_{\odot}$ Mpc$^{-3}$, the upper right panel
 shows the SFR density.  Our best-fit
model is also shown in each panel. Values of $\lambda_0=0.3$
and $\rho_{\rm BH,0}=2.5 \times 10^5 M_{\odot}$ Mpc$^{-3}$ 
and $\dot m_{\rm cr}=0$ are adopted here. Shaded areas represent
1-sigma confidence intervals of the model fits.
The lower left panel shows a direct comparison between best-fit normalized 
mass density of spheroids (solid line, red
shaded area) and black holes (dashed line, light red shaded
area). Also
shown is the evolution of the average Eddington scaled accretion rate
$\dot m(z)$ (dotted line, dark red shaded area). Finally, the lower right
panel shows a direct comparison between the best-fit 
SFR and BHAR (rescaled by a factor ${\cal A}_0
(1+\lambda_0)$) densities.}
\label{mer_fig:rhos0}       
\end{figure}

\section{SMBH as tracers of galaxy evolution}
Under the standard assumption that black holes grow mainly by
 accretion, their cosmic evolution 
can be calculated from the luminosity
 function of AGN $\phi(L_{\rm bol},z)=dN/dL_{\rm bol}$, where $L_{\rm
 bol}=\epsilon \dot M c^2$ is the bolometric luminosity produced by a
 SMBH accreting at a rate of $\dot M$ with a {\it radiative} efficiency
 $\epsilon$ (Soltan 1982). Following the discussion in MRD04, we will
 assume that the absorption corrected 2-10 keV luminosity
 function of AGN, $\phi(L_{\rm X},z)$, (La Franca et al. 2005) best describes the evolution of the {\it entire}
 accreting black holes population, yielding:
\begin{equation}
\label{mer_eq:rhobh_z}
\frac{\rho_{\rm BH}(z)}{\rho_{\rm BH,0}}=1-\int_0^{z}\frac{\Psi_{\rm BH}(z')}{\rho_{\rm BH,0}}\frac{dt}{dz'}dz',
\end{equation}
where the black hole accretion rate (BHAR) density is given by:
\begin{equation}
\label{mer_eq:bhar}
\Psi_{\rm BH}(z)=\int_0^{\infty}\frac{(1-\epsilon)L_{\rm bol}(L_{\rm
    X})}{\epsilon c^2}\phi(L_{\rm X},z)dL_{\rm X}
\end{equation}
$L_{\rm X}$ is the X-ray luminosity in the rest-frame 2-10 keV
band, and the bolometric correction function $L_{\rm bol}(L_{\rm X})$
is given by eq. (21) of Marconi et al. (2004). The exact
shape of $\rho_{\rm BH}(z)$ and $\Psi_{\rm BH}(z)$ then depends only on
the local black holes mass density $\rho_{\rm BH,0}$ and on
the (average) radiative efficiency $\epsilon$. This, in turn, is 
given by the product of the total accretion efficiency
$\eta(a)$, itself a function of the inner boundary condition 
and thus of the black hole spin parameter $a$, and a 
function $f$ of the Eddington scaled dimensionless accretion rate $\dot
m \equiv L_{\rm bol}/L_{\rm Edd}$. 
Below a critical rate, $\dot m_{\rm cr}$, accretion does not
proceed in the standard optically thick, geometrically thin
fashion for which $\epsilon$=$\eta$. Its radiative efficiency, instead, 
critically depends on the nature of
the flow: if powerful outflows/jets are capable of
removing the excess energy which is not radiated, as, for example, 
in the ADIOS scenario (Blandford \& Begelman 1999), then $f$=1, 
and black hole are always efficient
  radiators {\it with respect to the accreted mass} (``black holes are
  green!'', Blandford 2005). On the other hand, if advection across
  the event horizon is 
the dominant process by which energy is disposed of (ADAF, Narayan \& Yi 1995), we have:  
\begin{equation}
\label{mer_eq:eff}
\epsilon\equiv\epsilon(a,\dot m,\dot m_{\rm cr})=\eta(a) f(\dot m,
\dot m_{\rm cr})= \eta(a) \left\{
        \begin{array}{ll}
        1,   & \hbox{ $\dot{m} \ge \dot m_{\rm cr}$}   \\
        \dot{m}/\dot m_{\rm cr},   & \hbox{ $\dot{m} < \dot m_{\rm cr}$}   \\
        \end{array}\right.\;
\end{equation}
We use the total BHAR and mass densities,
$\Psi_{\rm BH}(z)$ and $\rho_{\rm BH}(z)$ respectively, to estimate the
redshift evolution of the average global Eddington scaled accretion rate
$\dot m(z) \propto \Psi_{\rm BH}(z)/\rho_{\rm BH}(z)$ (lower left
panel of Fig.~\ref{mer_fig:rhos0}) and the corresponding radiative
efficiency according to eq.~(\ref{mer_eq:eff}). This allows
us to identify the redshift at which 
a transition occurs in the global accretion
mode of growing SMBH. Depending on the assumed value of $\dot m_{\rm
  cr}$, 
this transition redshift is 0, if black holes are always efficient accretors,
i.e. $\dot m_{\rm cr}=0$, or $z\approx 0.6$ if $\dot m_{\rm cr}\approx 0.05$. 
Given the overall evolution of the BHAR density, 
this also implies that radiatively
inefficient accretion could contribute to only 
a small fraction of the total black hole mass
density (see also Yu and Tremaine 2002; Merloni 2004; Hopkins et al. 2006 and
references therein). 

\begin{figure}
\centering
\begin{tabular}{cc}
\includegraphics[height=5.cm]{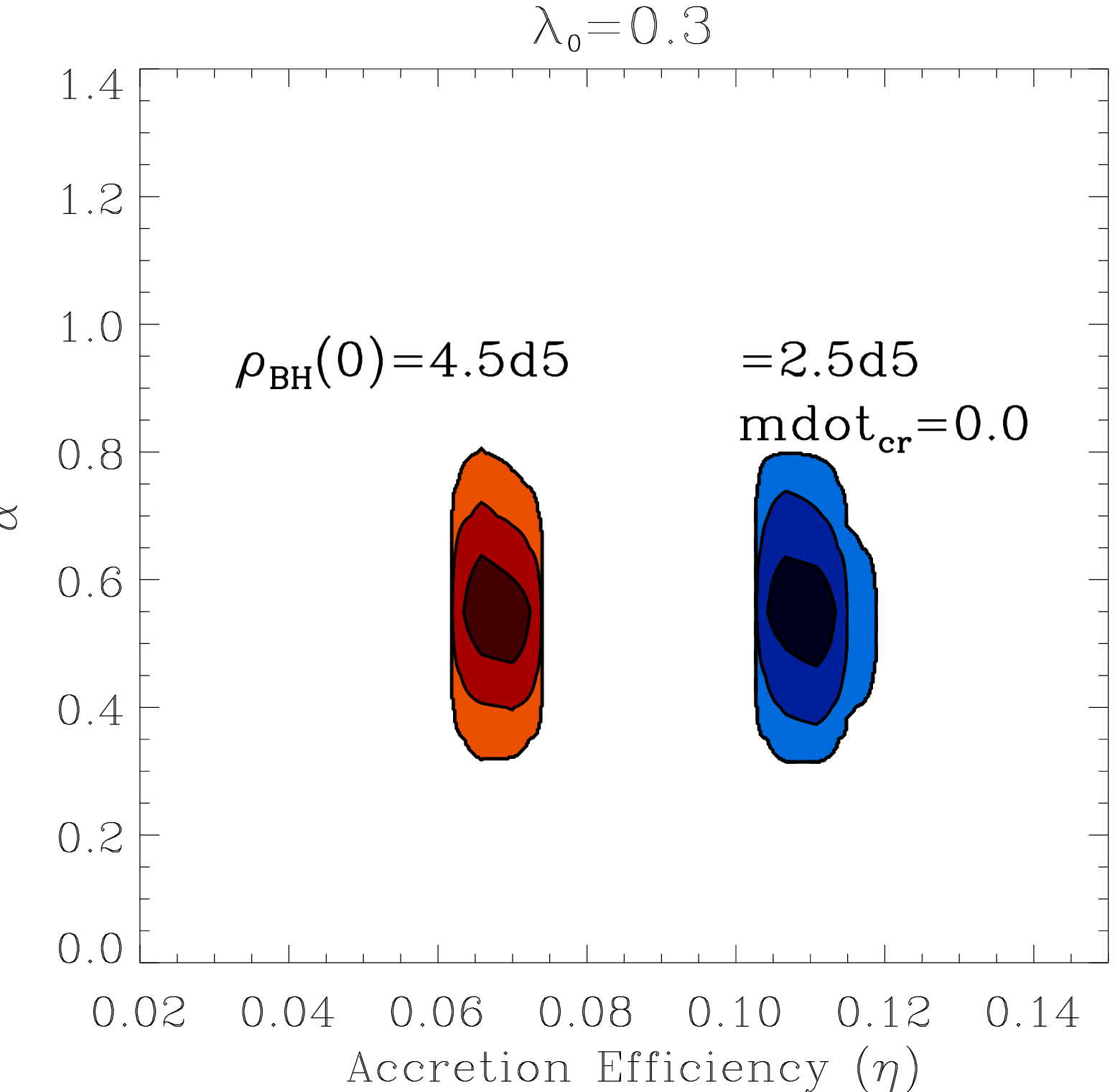} &
\includegraphics[height=5.cm]{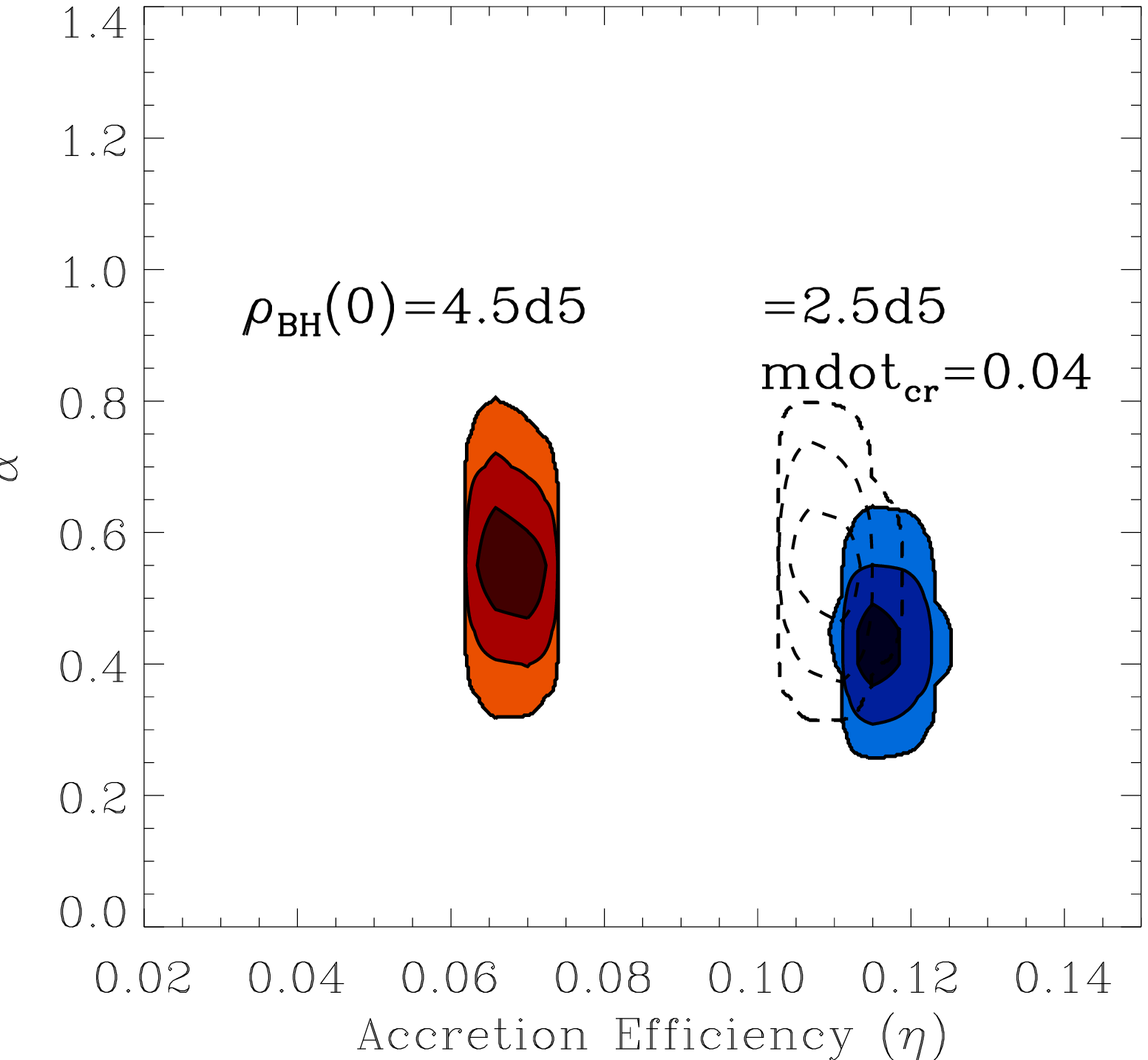} \\
\vspace{-0.3cm}
\end{tabular}
\caption{1,2 and 3 sigma confidence contours for the average accretion efficiency
  $\eta$ and the index $\alpha$ describing the evolution
  of the stellar spheroid to black hole mass density ratio. 
In each panel, the leftmost and rightmost
  set of contours show the results obtained assuming $\rho_{\rm BH,0}=4.5$
  (Marconi et al. 2004), and $2.5\times 10^{5}
  M_{\odot}$ Mpc$^{-3}$ (Yu and Tremaine 2002), respectively. Also shown
  is the dependence of such constraints on the assumed value of the
  critical rate between radiatively efficient and
  inefficient accretion (ADAF-like), with the dashed contours in the
  right panel showing the  $\dot m_{\rm cr}$=0 reference point.}
\label{mer_fig:ef_al}       
\end{figure}

\subsection{The parallel evolution}
Our goal is to link the growth of SMBH from 
eq.~(\ref{mer_eq:rhobh_z}) to the growth of stellar mass in galaxies.
Because local SMBH are observed to correlate with spheroids only, we
 introduce the parameter
 $\lambda(z)$, the ratio of the mass in disks and irregulars to that
 in spheroids at any redshift, so that the total stellar mass density can be
 expressed as: $\rho_{*}(z)=\rho_{\rm sph}(z)+\rho_{\rm disk+irr}(z)=\rho_{\rm
  sph}(z)[1+\lambda(z)]$. We then assume that $\lambda(z)$ 
evolves according to
  $\lambda(z)=\lambda_0 (1+z)^{-\beta}$, where $\lambda_0$
 is the value of the disk to spheroid ratio
 in the local universe. Also we assume that
the mass density of spheroids and supermassive black holes evolve in
parallel, modulo a factor $(1+z)^{-\alpha}$, obtaining a prediction for
the observable stellar mass density evolution as traced by SMBH growth:
\begin{equation}
\label{mer_eq:rhostar}
\rho_{*}(z)={\cal A}_{0}\rho_{\rm
  BH}(\epsilon,z)(1+z)^{-\alpha}[1+\lambda_0(1+z)^{-\beta}]
\end{equation}
where ${\cal A}_{0}$ is the constant of proportionality in the Magorrian
relation. By taking the derivative of (\ref{mer_eq:rhostar}), accounting for 
stellar mass loss,
an expression is also found for the 
corresponding star formation rate (SFR) density evolution 
(see eq. (7) of MRD04).

With these expressions  
we obtain statistically acceptable simultaneous fits to all 
available observational data points (see
MRD04 for a complete list of references) of both $\rho_*(z)$ and
SFR$(z)$. 
For each choice of 
$\rho_{\rm BH,0}$, $\lambda_0$, and of the critical accretion
rate $\dot m_{\rm cr}$, the fitting functions depend only on three
parameters: $\alpha$, $\beta$ and the accretion efficiency $\eta$. 
One example of such fits is shown in Fig.~\ref{mer_fig:rhos0} for the
specific case
$\rho_{\rm BH,0}=2.5\times 10^5 M_{\odot}$ Mpc$^{-3}$, $\lambda_0=0.3$
and  $\dot m_{\rm cr}=0$. Because the drop in the AGN integrated luminosity density
at low $z$ is apparently faster than that in SFR density (see lower right
panel of Fig.~\ref{mer_fig:rhos0}), the average black hole to spheroid mass
ratio must evolve with lookback time ($\alpha>0$; see lower left
panel of Fig.~\ref{mer_fig:rhos0}). This result is independent 
from the local black hole mass density, or from $\lambda_0$, 
and is not strongly affected by the choice of the value for the
critical accretion rate. 
This is shown also in Fig.~\ref{mer_fig:ef_al}, where the
constraints on the fit parameters are shown as confidence contours in the
accretion efficiency--$\alpha$ plane for various choices of $\dot
m_{\rm cr}$. As SMBH grow most of their mass
  at high $\dot m$, the global constraints on $\eta$ are not strongly
  affected by any reasonable choice of $\dot m_{\rm cr}$. 

\section{Conclusions}
We have made quantitative 
comparisons between the redshift evolution
of the integrated 
stellar mass in galaxies and the mass density of SMBH.  Although
clearly correlated with the stellar mass density, SMBH accretion does
not exactly
 track either the spheroid nor the total star assembly: 
irrespective of the exact mass budget in spheroids and
 disks~$+$~irregulars, the ratio of the total or spheroid stellar to
 black hole mass density was lower at higher redshift. 
 Our results also suggest that the fraction of stars locked up
 into the non-spheroidal components of galaxies and in irregular
 galaxies should increase with increasing redshift
 ($\beta<0$, see MRD04 for a discussion of this point). Our version of the Soltan
 argument yields a well defined constraint on the average radiative
 efficiency, and a corresponding one for the total accretion
 efficiency (i.e. on the mean mass-weighted SMBH spin), only weakly
 dependent on the uncertain physics of low $\dot m$ accretion flows.



\end{document}